\documentclass[%
 reprint,
 amsmath,amssymb,
 aps,
prb,
]{revtex4-2}

\usepackage{graphicx}
\usepackage{dcolumn}
\usepackage{bm}
\usepackage{hyperref}
\usepackage[mathlines]{lineno}
\usepackage{algorithm}
\usepackage[noend]{algpseudocode}
\usepackage{xcolor}
\usepackage{tikz}
\usetikzlibrary{matrix}

\def\pp{\texttt{++}}

\def\pq{\,\texttt{+=}\,}

\usepackage{soul}

\begin{document}
\preprint{arXiv}

\title{
  Fast Principal Minor Algorithms for Quantum Many Body Systems
}

\author{Fedor \v{S}imkovic IV$^{1,2}$}
\email{fsimkovic@gmail.com}
\author{Michel Ferrero$^{1,2}$}

\affiliation{
$^1$CPHT, CNRS, Ecole Polytechnique, Institut Polytechnique de Paris, Route de Saclay, 91128 Palaiseau, France\\
$^2$Coll\`ege de France, 11 place Marcelin Berthelot, 75005 Paris, France}


\date{\today}

\begin{abstract}
  The computation of determinants plays a central role in diagrammatic Monte Carlo algorithms for strongly correlated systems. The evaluation of large numbers of determinants can often be the limiting computational factor determining the number of attainable diagrammatic expansion orders. In this work we build upon the algorithm presented in [\emph{Linear Algebra and its applications} 419.1 (2006), 107-124] which computes all principal minors of a matrix in $O(2^n)$ operations. We present multiple generalizations of the algorithm to the efficient evaluation of certain subsets of all principal minors with immediate applications to Connected Determinant Diagrammatic Monte Carlo within the normal and symmetry-broken phases as well as Continuous-time Quantum Monte Carlo. Additionally, we improve the asymptotic scaling of diagrammatic Monte Carlo formulated in real-time to $O(2^n)$ and report speedups of up to a factor $25$ at computationally realistic expansion orders. We further show that all permanent-principal-minors, corresponding to sums of bosonic Feynman diagrams, can be computed in $O(3^n)$, thus encouraging the investigation of bosonic and mixed systems by means of diagrammatic Monte Carlo.
\end{abstract}

\maketitle

\section{Introduction} \label{intro}

Determinants are ubiquitous in the realm of mathematics and physics \cite{vein2006determinants, bjorklund2014determinant}. In particular, they play a crucial role in a variety of state-of-the-art numerical algorithms which compute the physical properties of fermionic many-body systems \cite{DQMC, rubtsov2005continuous, DDMC, cdet, rdet, rpadet, olivier}. 
Whilst a single determinant can be computed in polynomial time \cite{vein2006determinants}, many algorithms require the computation of a large number of principal minors\footnote{These are determinants of sub-matrices, obtained by removing a certain number of rows and corresponding columns.} of a given matrix.
In most cases, this results in exponential asymptotic scaling and often represents the computational bottleneck of the respective algorithms.

Diagrammatic Monte Carlo algorithms are based on the idea of expressing physical properties of interest as a perturbation series that is evaluated stochastically. A great asset of such algorithms is that they can directly treat systems in the thermodynamic limit (infinite system size) by computing exclusively connected Feynman diagrams. In early algorithmic implementations \cite{ProkofevSvistunovPolaronShort, kozik2010diagrammatic}, such connected Feynman diagrams were sampled individually in order to compute the coefficients of the relevant perturbation series.
Despite many successful applications \cite{kris_felix, deng, kun_chen, taheridehkordi2019algorithmic, vucicevic2019real}, these algorithms eventually suffer from the infamous fermion sign problem, which leads to increased statistical variance as a result of a
factorially growing number of sign-alternating diagrams with increasing expansion order. This difficulty has been overcome by a number of algorithms that efficiently compute sums of factorial numbers of diagrams by grouping them into determinants. 

Determinants have first made their way into diagrammatic Monte Carlo methods in the continuous-time Quantum Monte Carlo algorithms (CTQMC) \cite{rubtsov2005continuous,ctqmc_rmp}
and the related Determinant Diagrammatic Monte Carlo (DDMC) \cite{DDMC}. In these algorithms of polynomial complexity, the use of determinants strongly reduces the fermion sign problem by explicitly computing the sum of all connected and disconnected Feynman diagrams up to very high expansion orders ($\sim 1000$). CTQMC algorithms have proven especially useful in solving the quantum impurity models within dynamical mean-field theory~\cite{dmftreview, gullreview}, whilst DDMC has been successfully applied to study superfluid and magnetic transitions in the three-dimensional attractive \cite{DDMC} and repulsive models \cite{DDMC3DR} at half-filling. The presence of disconnected diagrams in the sums generated by determinants, however, has as consequence that the Monte Carlo variance grows exponentially with the number of orbitals or lattice sites. This leads to the necessity of extrapolating with system size in order to reach the thermodynamic limit and is another manifestation of the sign problem.

More recently, the Connected Determinant Diagrammatic Monte Carlo (CDet)\cite{cdet}
has been introduced. Just like the early diagrammatic Monte Carlo algorithms, it is formulated directly in the thermodynamic limit. However, instead of sampling individual diagrams, it evaluates the sum of the full factorial number of connected diagrams at a given expansion order at only exponential cost ($O(3^n)$) by recursively subtracting all disconnected diagrams from the sums generated with determinants. In order to be able to apply the recursion relations it is necessary to compute all principal minors of the originally evaluated matrix at each Monte Carlo step. A naive implementation of the principal minor computation scales as $O(n^3 2^n)$ and, despite being asymptotically favorable in comparison to the recursion step ($O(3^n)$), it constitutes the bottleneck at all realistically attainable expansion orders $n\leq 15$ (corresponding to matrix size). In Ref.~\onlinecite{griffin2006principal1} an algorithm for the simultaneous computation of all principal minors of a given matrix was presented, scaling more favorably as $O(2^n)$. This fast principal minor algorithm (FPM) has been used in most recent publications based on CDet computations \cite{fedor_sigma, fedor_hf, kim_cdet, lenihan2020entropy, M7, CDMFT, wietek}.

Recently, the CDet algorithm was generalized to perturbative expansions around symmetry-broken starting points inside of superfluid phases \cite{spada2021highorder}, using the Nambu formalism. For such computations, the need arises to compute a certain subset of 
$2^n$ principal minors of a $2n \!\times\! 2n$ matrix.

A determinant-based diagrammatic Monte Carlo algorithm has also been introduced for the real-time evolution of quantum systems \cite{olivier, corentin, corentin2, moutenet2019cancellation, QQMC, QQMC2}. In these algorithms, the computation of a contribution at perturbation
order $n$ requires to compute the sum of $2^n$ determinants of
$n \! \times \! n$ matrices, corresponding to a sum over Keldysh indices. Current state-of-the-art implementations compute this
sum in $O(n^2 2^n)$, which defines their asymptotic scaling.

For Feynman diagrammatic computations in bosonic systems, the CDet recursion equally allows one to extract all connected diagrams from sums generated by the evaluation of matrix permanents. Such an approach would be viable provided
the permanents of all principal sub-matrices can been obtained with reasonable computational effort.
However, the computation of a permanent scales exponentially as $O(n 2^n)$ \cite{ryser, glynn}, compared to the polynomial $O(n^3)$ scaling for determinants. This begs the question of whether the application of CDet to bosonic systems is computationally feasible and how such algorithms scale asymptotically.


In this paper, we introduce multiple generalizations of the FPM algorithm with useful applications to Diagrammatic Monte Carlo implementations, as described above. In particular, we show that the computational scaling of the real-time diagrammatic Monte Carlo within the Keldysh formalism can be improved to $O(2^n)$. We further show that the same scaling also applies to the computation of principal minors in CDet within symmetry-broken phases. We also present an $O(3^n)$ CDet generalization to diagrammatic expansions that involve two vertex types. Finally, we show that summing all connected diagrams for bosonic and mixed (Fermi-Bose) systems can have the same computational scaling as for purely fermionic ones, $O(3^n)$.

The paper is structured as follows: In Section~\ref{principalminors} we introduce the original Fast Principal Minor algorithm as well as its generalizations in Section~\ref{generalisations} and introduce a principal minor algorithm for permanents in Section~\ref{permanents}. In Section~\ref{applications} we provide details of specific diagrammatic Monte Carlo application, before discussing conclusions in Section \ref{conclusions}.

\section{Fast Principal Minors algorithm} \label{principalminors}

Our aim is to compute all principal minors of a given $n\!\times\! n$ square matrix $A[\mathcal{S}]$, where $\mathcal{S} = \{1,2, \dots, n \}$ is a set of rows and columns. A principal minor $A[\mathcal{S}_i]$, corresponding to a subset $\mathcal{S}_i \subseteq \mathcal{S}$, is the determinant of
the sub-matrix generated by keeping only the rows and columns specified by $\mathcal{S}_i$.
We denote by $\bar{\mathcal{S}}_i$ the complement of $\mathcal{S}_i$ with respect to the
full set $\bar{\mathcal{S}}_i \equiv \mathcal{S}/\mathcal{S}_i$.
We further denote non-square matrices as $A[\mathcal{S}_i,\mathcal{S}_j]$ and we
have $A[\mathcal{S},\mathcal{S}] \equiv A[\mathcal{S}]$. Following standard practice, the determinant of an empty matrix is defined as $\det(A[\emptyset]) = 1$.
In the following, we will always associate $\mathcal{S}_i \subseteq \mathcal{S}$ with the subset described by the binary bit-mask provided by the integer
$i$ (e.g. $\mathcal{S}_{13} = \{1,3,4\}$).

A trivial way to compute all principal minors is to generate each $k\times k$ sized sub-matrix individually and simply compute the determinant by means of an algorithm with polynomial $\mathcal{O}(k^3)$ complexity, e.g. Gaussian elimination or LU decomposition \cite{trefethen1997numerical}. The total number of operations needed to compute all principal minors this way
scales exponentially as $\mathcal{O}(n^3 2^n)$ and requires polynomial space ($O(n^2)$).

This computational scaling can often be reduced to $\mathcal{O}(n^2 2^n)$ by making use of Sherman–Morrison–Woodbury type-formulas \cite{sherman1950adjustment} and changing determinants by one row and column at a time using a Gray code.

A computationally superior algorithm scaling as $O(2^n)$ was presented in Ref.~\cite{griffin2006principal1}. Below we provide the main ideas of the algorithm without the mathematically rigorous derivations that can be found in the original publication and references therein \cite{crabtree1969identity,horn2012matrix,tsatsomeros2000recursive,li2009asymptotic}.

We start from the well known relation \cite{horn2012matrix}:
\begin{align}
	\det(A[\mathcal{S}]) = \det(A[\mathcal{S}_i])\det(A[\mathcal{S}]/A[\mathcal{S}_i]),
    \label{dets_schur}
\end{align}
where we have introduced the Schur complement, which we define as:
\begin{align}
A[\mathcal{S}]/A[\mathcal{S}_i] \equiv A[\bar{\mathcal{S}}_i] - A[\bar{\mathcal{S}}_i,\mathcal{S}_i] A[\mathcal{S}_i]^{-1} A[\mathcal{S}_i,\bar{\mathcal{S}}_i].
\label{eq:schur}
\end{align}

In order to proceed we need two identities. It should be noted that both relations only hold for non-singular matrices and non-singular principal minors thereof.
The first identity shows that eliminating selected rows and columns can be done either before or after taking the Schur complement
without affecting the computation of the determinant:
\begin{align}
\det(A[\mathcal{S}_j \cup \mathcal{S}_k]) = \det(A[\mathcal{S}_j])\det((A[\mathcal{S}]/A[\mathcal{S}_j])[\mathcal{S}_k]),
\label{lemma1}
\end{align}
where $\mathcal{S}_k \subseteq \bar{\mathcal{S}}_j$.
The second identity, called the quotient property, is necessary to treat nested Schur complementation:
\begin{align}
A[\mathcal{S}]/A[\mathcal{S}_j \cup \mathcal{S}_k] = (A[\mathcal{S}]/A[\mathcal{S}_j])/((A[\mathcal{S}]/A[\mathcal{S}_j])[\mathcal{S}_k]).
\label{lemma2}
\end{align}
This means that instead of taking one Schur complement with respect to a large sub-matrix one can successively take multiple Schur complements with respect to single matrix elements. 
\begin{figure}
\centering
\includegraphics[width=0.5\textwidth]{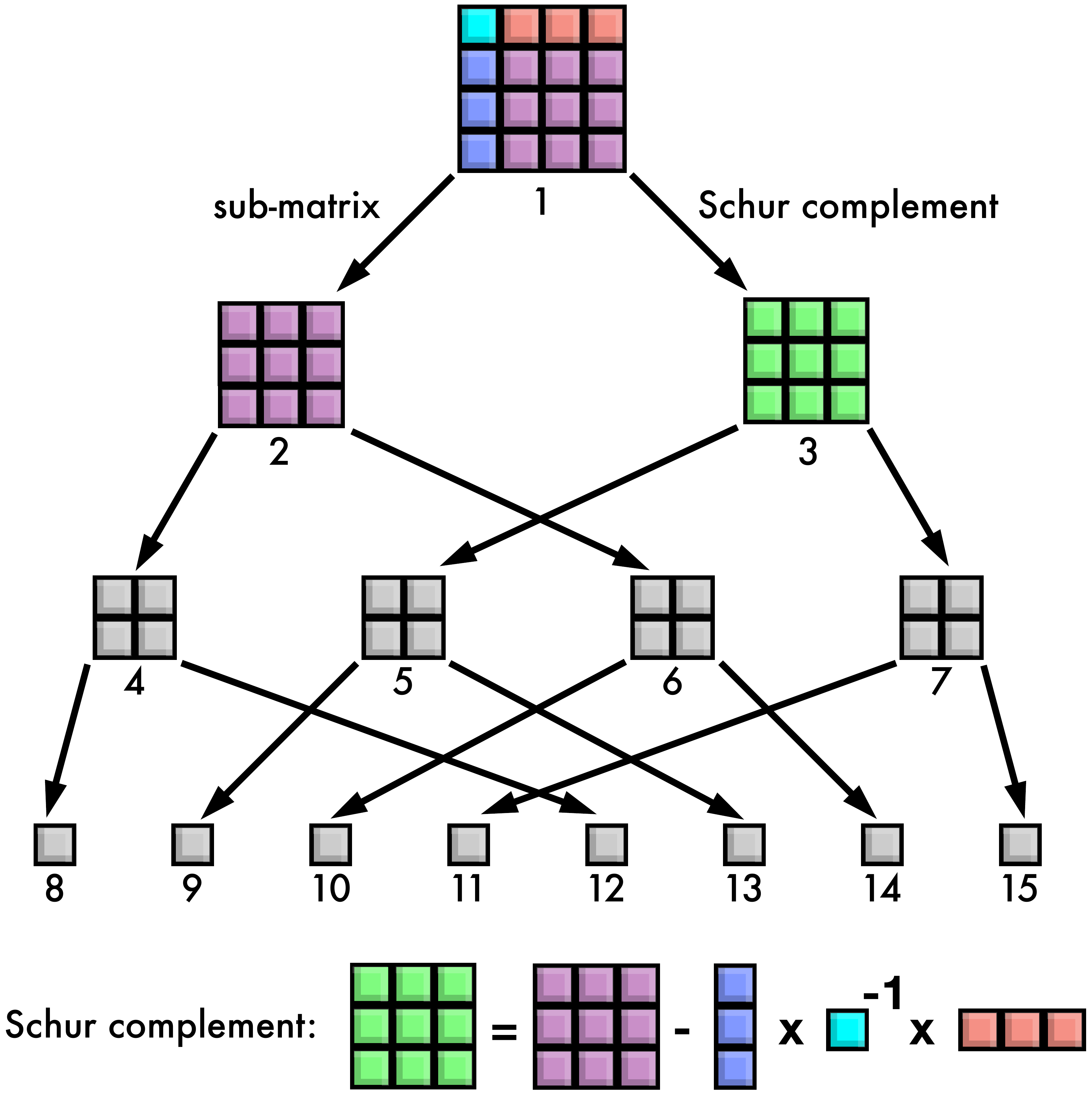}
\caption{The binary tree corresponding to the algorithm for the computation of principal minors of a matrix of size $4\!\times\!4$.
From each node, the sub-matrix with the first row and column removed is copied along the left branch. The Schur complement
with respect to the $(1,1)$ element is taken along the right branch.}
\label{Fig:minors_tree}
\end{figure}

The algorithm presented in Ref.~\onlinecite{griffin2006principal1} generates a binary tree with $2^n - 1$ nodes
that are used to compute the different principal minors of the matrix $A[\mathcal{S}]$. Each node is labeled
with an integer, beginning with $1$ at the first level
and enumerating the following nodes in a top to bottom and left to right
order as portrayed in Fig.~\ref{Fig:minors_tree}. A matrix is associated with every node of the tree, starting
with the original full matrix $A[\mathcal{S}]\equiv A^{(1)}$ for the first node $i=1$ at level $l=1$.
Whenever we take the left branch from a node $i$ at level $l$ we generate a new sub-matrix
$A^{(i+2^{l-1})}$ at the node $(i+2^{l-1})$ by removing the first row and column of $A^{(i)}$. Whenever we take the right branch
we generate a matrix $A^{(i+2^l)}$ at the node $(i+2^l)$ from the Schur complement of $A^{(i)}$ with respect to its $(1,1)$ element:
\begin{align}
  &A^{(i+2^{l-1})} = A^{(i)}[\{l+1,\dots,n\}] \quad \quad \text{left branch}\\
  &A^{(i+2^{l})\phantom{^{+1}}} = A^{(i)}/A^{(i)}[\{l\}] \quad \quad \quad \quad \; \, \text{right branch}
\end{align}
As can be seen from the procedure above, the integer that labels a node in the tree can be interpreted as
a binary mask keeping track of which columns and rows have been retained or eliminated before reaching that node.
One can then compute the principal minor associated with the node $i$
at level $l$ by using Eq.~\ref{lemma1}:
\begin{align}
\det(A[\mathcal{S}_i]) &\equiv \det(A[\mathcal{S}_j \cup \{l\}]) \nonumber \\ &= \det(A[\mathcal{S}_j]) \, \det(A[\mathcal{S}]/A[\mathcal{S}_j])[\{l\}] \nonumber \\
   &= \det(A[\mathcal{S}_j]) \, A_{1,1}^{(i)},
\label{eq:minor}
\end{align}
where we have introduced $\mathcal{S}_j \equiv \mathcal{S}_i / \{l\}$.
Applying Eq.~\ref{eq:minor} to all nodes of the tree eventually yields all principal minors
of the matrix $A[\mathcal{S}]$.



Sometimes, there is one additional difficulty that needs to be taken into account.
Whenever the Schur complement is taken (as per Eq.~\ref{eq:schur}) it is necessary to calculate the inverse of a matrix element called the pivot. If the pivot's value is zero or vanishingly small this procedure forcibly results in numerical instabilities of the algorithm. Ref.~\cite{griffin2006principal1} provides a solution for such cases by adding a correction term $C$ to the pivot which must be, however, compensated for at the end of the computation. Any time a pivot is changed this affects all principal minors computed from the resulting Schur complement. After all principal minors have been computed a pivot-correction has to be back-propagated through principal minors for each instance where the correction term was added.

For example, let $\{l\}$ be a very small (or vanishing) element corresponding to the first row and column of the matrix $A[\mathcal{S}_i]$.
To avoid numerical instabilities, we change $A[\mathcal{S}_i]_{1,1} \rightarrow A[\mathcal{S}_i]_{1,1} + C$ and use this modified
matrix $\tilde{A}[\mathcal{S}_i]$ to compute the Schur complement. In order to recover the
determinant of the original matrix $A[\mathcal{S}_i]$ one can use the formula:
\begin{align}
  \det(A[\mathcal{S}_i]) = \det(\tilde{A}[\mathcal{S}_i]) - C  \det(A[\mathcal{S}_i / \{l\}]),
\end{align}
which is how the pivot-correction is back-propagated at the end of the computation. The minimal number of pivot-corrections that need to be performed for an $n\!\times\! n$ matrix with all diagonal entries being zero, as is often the case in diagrammatic Monte Carlo algorithms \cite{rubtsov2005continuous, DDMC, cdet}, is $n-1$, which results in additional computational cost of $O(2^n)$ for the back-propagation loop. If a pivot-correction would be performed at every instance of Schur complementation the resulting computational cost would become $O(n 2^n)$ instead.

Below is a sketch of the FPM algorithm:
\begin{algorithm}[H]
\caption{Fast Principal Minor Algorithm}\label{algFPM}
\begin{algorithmic}[1]
\For{$l=1\quad l\leq n \quad l\pp$}
\For{$i=2^{l-1} \quad i<2^l\quad i\pp$}
\State $\det(A[\mathcal{S}_i])\gets   \det(A[\mathcal{S}_i / \{l\}]) \, A_{1,1}^{(i)}$
\State $A^{(i+2^{l-1})} \gets A^{(i)}[ \{l+1,\dots,n\}]$  \Comment{left branch}
\State $A^{(i)}_{1,1} \pq C$ \Comment{pivot correction (optional)}
\State $A^{(i+2^{l})} \gets A^{(i)}/A^{(i)}[\{l\}]$ \Comment{right branch}
\EndFor
\EndFor
\end{algorithmic}
\end{algorithm}


The computational scaling of the algorithm is easily found to be $\sum_{l=0}^{n-2} 2^l \cdot 2(n-l-1)^2 \approx O(2^n)$ operations.  Ref.~\onlinecite{griffin2006principal1} provides a breadth-first-traversal algorithm to calculate all principal minors, as summarized in Alg.~\ref{algFPM}. In this formulation all matrices computed at a particular level of the tree need to be stored, which leads to exponential memory costs of $O(2^n)$. It is possible, however, to traverse the tree depth-first or formulate a recursive version of the algorithm in order to reduce the scaling to the polynomial $O(n^3)$ as only one matrix needs to be stored at each level. This can lead to significant accelerations for larger matrices, overcoming the slowdown reported in Ref.~\cite{griffin2006principal1} (see Fig.\ref{Fig:CDet1}).

\section{Generalizations to Principal Minor Subsets}  \label{generalisations}

In this subsection we want to address cases, when it is desired to compute only a certain subset of all principal minors of a matrix.

\subsection{Leading principal minors}

It can be straightforwardly seen that is it possible to use successive Schur-complementation to compute the principal minor corresponding to the determinant of the full $n\!\times\!n$ matrix in $O(n^3)$ operations. The disadvantage with respect to other more established algorithms, such as Gaussian elimination, is that there is no direct way to implement pivot corrections in the way it is done for the FPM algorithm. This is due to the necessity of subtracting principal minors at the back-propagation step, which have not been computed in the process. On the other hand, this algorithm has the advantage that it computes all \emph{leading} principal minors corresponding to subsets corresponding to indices $2^l-1$ with $1\leq l\leq n$ in the process.

\subsection{XNOR principal minors of a $2n \!\times\! 2n$ matrix}

\begin{figure}
\centering
\includegraphics[width=0.3\textwidth]{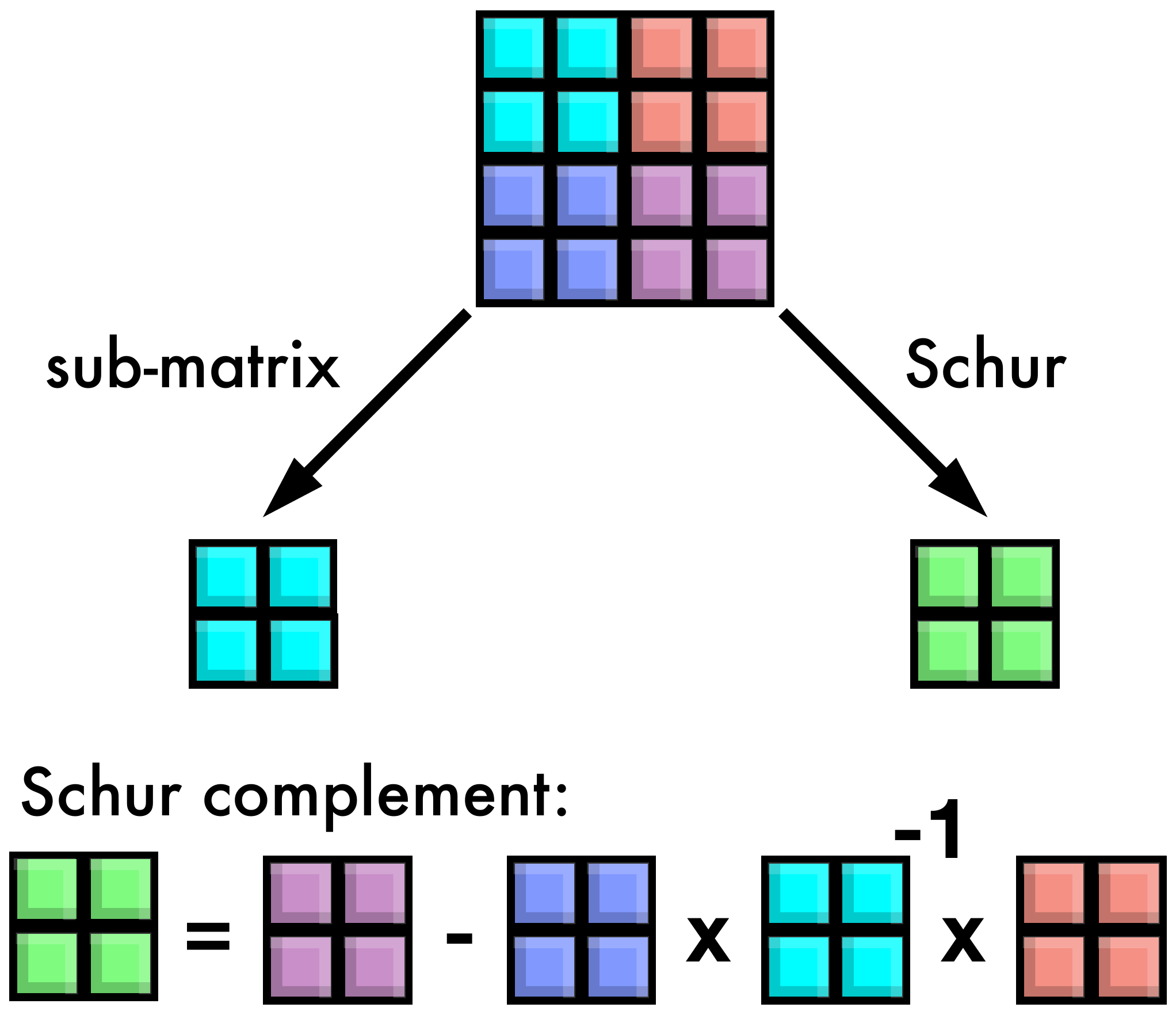}
\caption{Binary tree for used in the algorithm for the computation of principal minors of a matrix of size $4\!\times\!4$, corresponding to sub-matrices respecting the XNOR case rules.}
\label{Fig:FPM_2n}
\end{figure}

Now, let us consider a $2n \!\times\! 2n $ matrix $A[\mathcal{S}]$. Computing all principal minors of $A[\mathcal{S}]$ with the FPM algorithm takes $O(4^n)$ operations, proportional to the total number of principal minors, $4^n-1$. In some cases it is, however, sufficient to compute only a certain subset of minors. Specifically, if we treat $A[\mathcal{S}]$ as an $n \!\times\! n$ block-matrix with $2 \!\times\! 2$ blocks then some cases are of particular interest. The first one is: Compute principal minors of all subsets $\mathcal{S}_i$ such that for all $0\leq l < n$ we satisfy one of the following two options:
\begin{itemize}
  \item $\{2l+1\} \in \mathcal{S}_i$ and $\{2l+2\} \in \mathcal{S}_i$
  \item $\{2l+1\} \notin \mathcal{S}_i$ and $\{2l+2\} \notin \mathcal{S}_i$
\end{itemize}
We will call this set of options the \emph{XNOR case}.
In the language of matrices this means that either both the $(2l+1)$-th and $(2l+2)$-th rows and columns are included in a sub-matrix or none of the two. One must then evaluate a total of $2^n-1$ principal minors, which can be done naively in $O(n^3 2^n)$ by computing them separately. It is readily seen that one can form a binary tree with $2^n-1$ nodes, corresponding to the principal minors of interest, by removing two rows and columns from a matrix when taking the left branch and by performing the corresponding Schur complement whenever taking the right branch, see Fig.~\ref{Fig:FPM_2n}. This procedure computes all the relevant ratios of principal minors of sub-matrices adhering by the aforementioned rules. Since the pivot, of which we must compute the inverse, has become a $2\!\times\!2$ matrix and since we compute only a subset of the principal minors of the original matrix, it is no longer possible to add corrections to pivots. The algorithm scales as $O(2^n)$ and is roughly four times slower than FPM for an $n\!\times\!n$ matrix due to the fact that the Schur complement is taken with respect to $2\!\times\!2$ matrices instead of a single element.

It is possible to extend this idea to $k n \!\times\! k n$ sized matrices with \emph{steps} of size $k$ resulting in asymptotic scaling of $O(k^2 2^n)$.

\subsection{NAND principal minors of a $2n \!\times\! 2n$ matrix}

\begin{figure}
\centering
\includegraphics[width=0.3\textwidth]{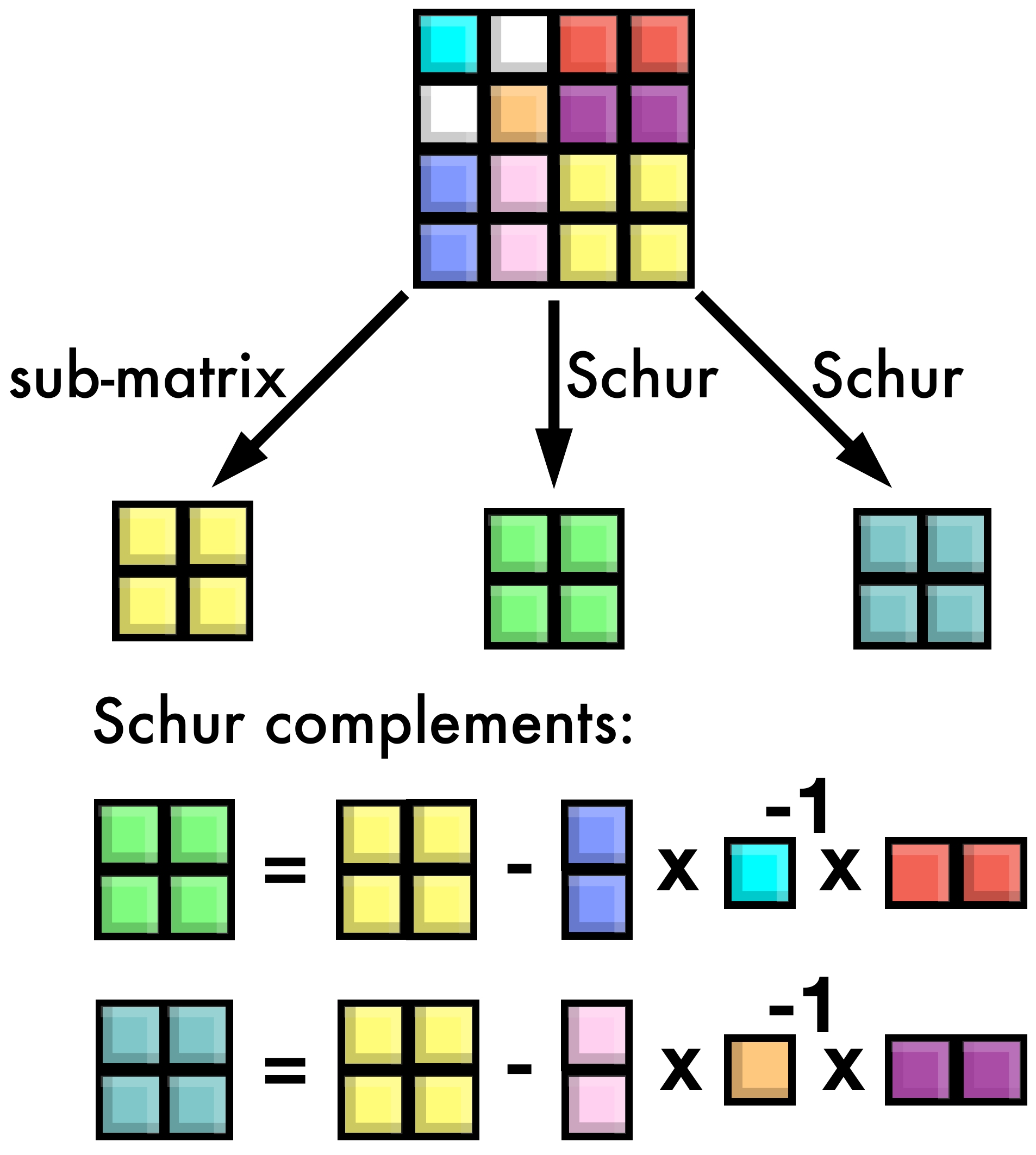}
\caption{Binary tree corresponding to the algorithm for the computation of principal minors of a matrix of size $4\!\times\!4$, corresponding to sub-matrices respecting the NAND case rules. Note that off-diagonal matrix $A[\mathcal{S}]_{i,j}$ elements with $|i-j|=1$ (white) do not ever enter the calculations.}
\label{Fig:FPM_3n}
\end{figure}

In the second case we are interested in computing all principal minors such that for all $0\leq l< n$ we satisfy one of the following three options:
\begin{itemize}
  \item $\{2l+1\} \! \in \mathcal{S}_i$ and $\{2l+2\} \! \notin \mathcal{S}_i $
  \item $\{2l+1\} \! \notin \mathcal{S}_i$ and $\{2l+2\} \! \in \mathcal{S}_i $
  \item $\{2l+1\} \! \notin \mathcal{S}_i$ and $\{2l+2\} \! \notin \mathcal{S}_i $
\end{itemize}
We will call this the \emph{NAND case}. In matrix terms this means that either the $(2l+1)$-th or the $(2l+2)$-th row and column are included in a sub-matrix of interest or none of the two. The total number of corresponding principal minors is $3^n-1$. Here, it is possible to transform the original $2n\!\times\!2n$-sized block-matrix $A[\mathcal{S}]$ into an $n\!\times\!n$-sized matrix $\hat{A}[\mathcal{S}]$ with polynomial entries in $\alpha_i^{\pm}$ such that:
\begin{align}
  \hat{A}_{i,j} &= A_{2i+1,2j+1} \, \alpha_i^+ \alpha_j^+ + A_{2i+2,2j+1}\,  \alpha_i^- \alpha_j^+ \\&+ A_{2i+1,2j+2}\,  \alpha_i^+ \alpha_j^- + A_{2i+2,2j+2}\,  \alpha_i^- \alpha_j^- \nonumber
\end{align}
with the aforementioned rules determining which principal minors need to be evaluated being encoded by:
\begin{align}
  \alpha_i^+ \alpha_i^- = 0.
\end{align}
One can then proceed to evaluate the $n\!\times\!n$ polynomial-valued matrix $\hat{A}[\mathcal{S}]$ with the FPM algorithm. This means that pivot corrections can be added in the same way as they would be in the FPM algorithm, as long as they are back-propagated at the end of the calculation. An alternative way of looking at the new representation is as the rearrangement of $3^n-1$ relevant nodes from the original binary tree of depth $2n$ into a ternary tree of depth $n$, see Fig.\ref{Fig:FPM_3n}. Then, the \emph{left} branch is obtained by copying a sub-matrix eliminating the first two rows and columns. Two additional, \emph{right}, branches are obtained as the result of taking the Schur complement with respect to the $(1,1)$ and $(2,2)$ elements of the matrix at the given node. The total computational cost of the algorithm is governed by the number of Schur complements taken at each level of the ternary tree $ \sum_{l=0}^{n-2} 3^l \cdot 4(n-l-1)^2 \approx O(3^n)$. 

It is straightforward to extend this algorithm to $kn \times kn$ sized matrices resulting in asymptotic scaling of $O((k+1)^n)$, related to the the number of relevant principal minors: $(k+1)^n-1$.

\subsection{XOR principal minors of a $2n \!\times\! 2n$ matrix}

In the third case we are interested in computing all principal minors such that for all $0\leq l< n$ we satisfy one of the following two options:
\begin{itemize}
  \item $\{2l+1\} \! \in \mathcal{S}_i$ and $\{2l+2\} \! \notin \mathcal{S}_i $
  \item $\{2l+1\} \! \notin \mathcal{S}_i$ and $\{2l+2\} \! \in \mathcal{S}_i $
\end{itemize}
We will call this the \emph{XOR case}. This is equivalent to only computing principal minors of $n \!\times\!n$ sized sub-matrices of $\hat{A}[\mathcal{S}]$ and adhering by the rules of the NAND case. One can resort to evaluating only the two \emph{right}, Schur-complement branches in the ternary tree of Fig.~\ref{Fig:FPM_3n}, which effectively reduces it to a binary tree. This algorithm can be shown to lead to a reduced scaling of $O(2^n)$. As in the case of computing a single determinant using Schur complements, this has the advantage of computing lower order principal minors as a side-product. The downside, similarly to the single determinant case, is that this algorithm doesn't allow for pivot corrections without significant additional computational cost. 

Similarly to the previous two cases, one can generalize this algorithm to $k n \!\times\! k n$ sized matrices resulting in asymptotic scaling of $O(k^n)$.

\section{Permanent Principal Minor Algorithm}\label{permanents}

Whilst the determinant of an $n \!\times\! n$ sized matrix can be computed in polynomial time $O(n^3)$, the two most efficient algorithms to compute the permanent of the same matrix, found by Ryser and Glynn, scale exponentially as $O(n 2^n)$ \cite{ryser, glynn, valiant}. The inclusion-exclusion Ryser formula for the computation of a permanent reads:
\begin{align}
\operatorname{per}(A) = \sum_{\mathcal{S}_k \subseteq \{1,\dots,n\} } \!\! \left(-1\right)^{|\mathcal{S}_k|} \prod_{i=1}^n \sum_{j \in \mathcal{S}_k} A_{i,j}.
\label{ryser}
\end{align}
A numerical implementation hereof consists of two steps. First, one precomputes all possible sums over index $j$ given by $\sum_{j \in \mathcal{S}_k} A_{i,j}$ in Gray code order with $O(n 2^n)$ additions. Subsequently, one evaluates the products over sums for all $2^n-1$ subsets $\mathcal{S}_k \subseteq \{1,\dots,n\}$ using $O(n 2^n)$ multiplications which defines the asymptotic scaling of the algorithm. As a consequence, a naive successive application of the Ryser algorithm to compute all permanent principal minors of a matrix results in a scaling of $O(n 3^n)$.

We can, however, reduce this scaling by generalizing the algorithm to the case when all permanent principal minors of a matrix are desired. Let us  rewrite the formula of Eq.~\ref{ryser} for a given subset (minor) $\mathcal{S} \subseteq \{1,\dots,n\}$ as:
\begin{align}
\operatorname{per}(A[\mathcal{S}]) = \sum_{\mathcal{S}_k \subseteq \mathcal{S} } \!\! \left(-1\right)^{|\mathcal{S}_k|} \prod_{i \in \mathcal{S}} \sum_{j \in \mathcal{S}_k} A_{i,j}
\end{align}
Here, the first (additions) step remains identical as all possible sums are already being computed in the computation of the full matrix permanent. For the second step it is necessary to evaluate all products corresponding to subsets $\mathcal{S}$, given as:
\begin{align}
  \Pi[\mathcal{S}] = \prod_{i \in \mathcal{S}} \sum_{j \in \mathcal{S}_k} A_{i,j}
\end{align}
These subsets can be evaluated in Gray code order, which, however, entails the necessity of using divisions. A division-free way of evaluating subsets can be achieved by using a recursion relation:
\begin{align}
  \Pi[\mathcal{S}] = \Pi[\mathcal{S}/ \{j\}] \sum_{i \in \mathcal{S}_k} A_{i,j}
\end{align}
where $j \in \mathcal{S}$ can be chosen arbitrarily. This corresponds to $\sum_{k=1}^n \binom{n}{j} 2^k = O(3^n)$ operations, which sets the asymptotic scaling of the algorithm. Remarkably, the step resembles a subset convolutions of two sets, which is essentially what is being computed when applying the recursive relation within the CDet algorithm \cite{cdet}. This means that a fast subset convolution algorithm \cite{koivisto} is applicable, which would reduce the asymptotic scaling further, to $O(n^2 2^n)$, albeit that only becomes practically advantageous for matrix-sizes of $n \gtrsim 15$.

Interestingly, a related inclusion-exclusion principle based fast diagram summation algorithm has been found for strong-coupling expansions of fermionic \cite{boag} as well as bosonic \cite{yang2021inclusion} systems. 


\section{Applications to diagrammatic Monte Carlo algorithms} \label{applications}

Let us now focus on applications to various numerical algorithms for quantum many body systems based on diagrammatic Monte Carlo. In such algorithms one is generally interested in computing a physical observable $\mathcal{C}$ expressed in terms of a power series:
\begin{align}
  \mathcal{C}(\xi) = \sum_{n=0}^{\infty} a_n \xi^n
  \label{eq:series}
\end{align}
where the coefficients $a_n$ are obtained from the stochastic evaluation of Feynman diagrams, integrated over internal space and imaginary-time variables $X = (\vec{r},\tau)$.

\subsection{CDet}

\begin{figure}
\centering
\includegraphics[width=0.48\textwidth]{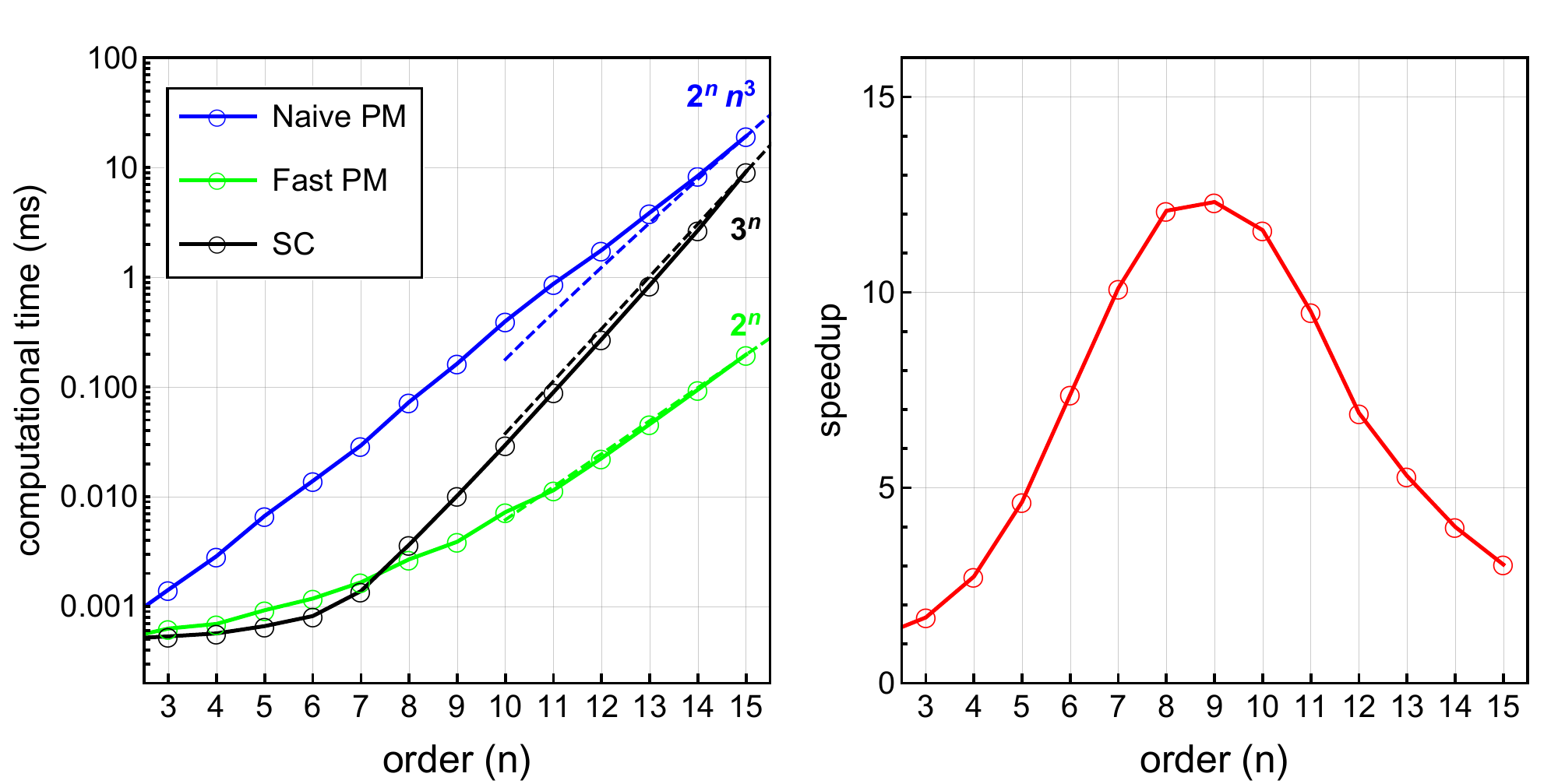}
\caption{\emph{Left}: Computational time as a function of order ($n$) for the fast principal minor (PM) algorithm as compared to the naive determinantal implementation as well the subset convolution (SC) used by CDet. \emph{Right}: Computational speedup of the complete CDet algorithm using fast principal minors with respect to the algorithm using a naive determinant implementation.}
\label{Fig:CDet1}
\end{figure}

The Fast Principle Minor algorithm has first been used in the context of the
CDet \cite{cdet} algorithm. In this method, the contribution $c[\mathcal{S}]$ to a
coefficient $a_n$ in~Eq.~\ref{eq:series}
for a given set $\mathcal{S} = \{ X_1, \ldots, X_n \}$
of internal vertices is the sum of all connected diagrams constructed on $\mathcal{S}$.
This sum is obtained by a recursion formula that involves the prior computation of all
principal minors of a given matrix corresponding to subsets
$\mathcal{S}_i \subseteq \mathcal{S}$:
\begin{align}
  a[\mathcal{S}_i] = \det(M_n^{\uparrow}[\mathcal{S}_i]) \det(M_n^{\downarrow}[\mathcal{S}_i]),
\end{align}
where $\left(M_n^{\sigma}\right)_{i,j} = G_{0 \sigma}(X_i-X_j)$.
Then, a recursive relation is used to compute sums of connected diagrams $c[\mathcal{S}]$ from $a[\mathcal{S}]$ by subtracting all disconnected ones:
\begin{align}
  c[\mathcal{S}] = a[\mathcal{S}] - \sum_{\substack{\mathcal{S}_i \subsetneq \mathcal{S} \\ \{j\} \in \mathcal{S}_i}} c[\mathcal{S}_i] a[\mathcal{S}/\mathcal{S}_i].
\label{eq:recursion}
\end{align}
where $\{j\} \in \mathcal{S}$ is chosen arbitrarily. The recursive algorithm corresponds to a subset convolution and can be evaluated in $O(3^n)$ computational steps, or alternatively in $O(n^2 2^n)$ by using fast subset convolutions \cite{koivisto} which, however, in practice performs slightly worse for attainable orders $n \lesssim 15$. The total computational time of the algorithm is thus limited by both the computation of principal minors and the subset convolution, depending on the expansion order, see Fig.~\ref{Fig:CDet1}. We see that the FPM algorithm significantly outperforms a naive determinant implementation and becomes sub-leading to the subset convolution around order $n=8$. The total speedup of the CDet algorithm due to the use of FPM also peaks at a factor $12$ around orders $8 \lesssim n \lesssim 10$, which corresponds to the highest attainable orders for calculations in challenging regimes of the Hubbard model. At yet higher orders the gain reduces due to an ever increasing contribution of the subset convolution to the total computational time.

We also note that at order $24$ our $\text{C}\texttt{++}$ implementation of FPM algorithm takes $0.23$ seconds on a standard laptop, compared to $443$ seconds reported in Ref.~\cite{griffin2006principal1}, roughly corresponding to an improvement by a factor $2000$.

\begin{figure}
\centering
\includegraphics[width=0.48\textwidth]{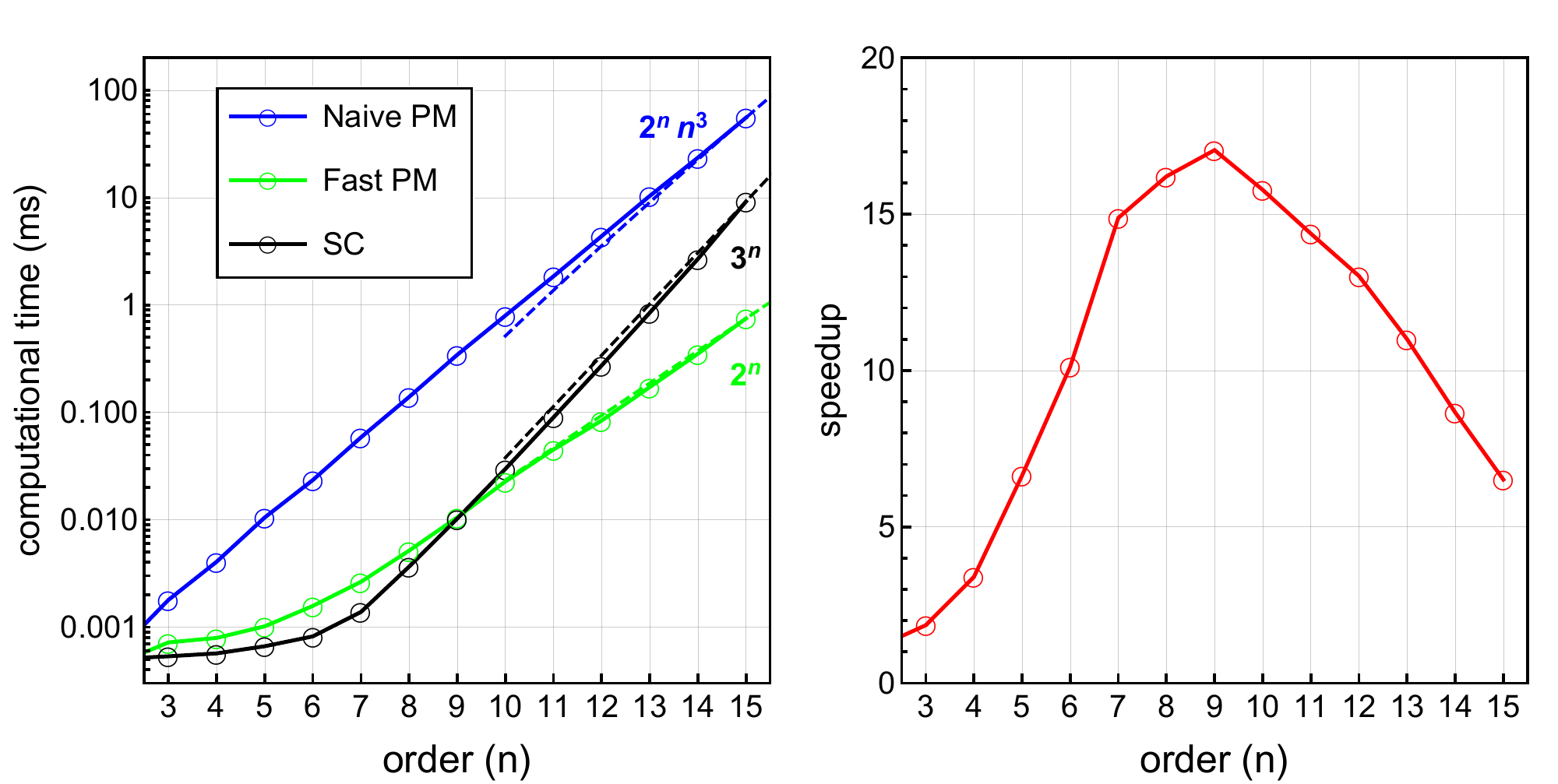}
\caption{\emph{Left}: Computational time as a function of order ($n$) for the fast principal minor (PM) algorithm within CDet in the Nambu formalism as compared to the naive determinantal implementation as well the subset convolution (SC) used by CDet. \emph{Right}: Computational speedup of the complete Nambu CDet algorithm using fast principal minors with respect to the algorithm using a naive determinant implementation.}
\label{Fig:CDet2}
\end{figure}

\subsection{CDet for expansions around BCS theory}

In a recent work \cite{spada2021highorder} the CDet algorithm was generalized to expansions around
the BCS mean-field theory inside superfluid phases and using the Nambu formalism \cite{BCS}. The main algorithmic difference to CDet in the paramagnetic regime is that one needs to evaluate $2n\!\times\!2n$ sized block-matrices with bare Nambu propagators as entries. These matrices
are built from $2\!\times\!2$ blocks $\tilde{M}_{i,j}$ given by:
\begin{align}
  \tilde{M}_{i,j}  =
  \begin{pmatrix}
  G_{00}(X_{ij}) & G_{01}(X_{ij}) \\
  G_{10}(X_{ij}) & G_{11}(X_{ij})
  \end{pmatrix}
\end{align}
with $i,j \in \{1, \ldots, n\}$ and $X_{ij} \equiv X_i-X_j$. It is necessary to compute $2^n-1$ principal minors of this matrix, exactly corresponding to the XNOR case from the previous section. The recursion step involving a subset convolution does not differ from the original CDet one. From Fig.\ref{Fig:CDet2} we see that, similarly to the original CDet algorithm, a FPM algorithm significantly outperforms the naive determinant implementation. The maximum computational gain of a factor $17$ is achieved at order $n=9$ before slowly decreasing for higher orders due to the computational cost of the subset convolution taking over. Again, the gain is maximal at orders which correspond to the limits of many realistic computations in difficult regimes.

\begin{figure}
\centering
\includegraphics[width=0.48\textwidth]{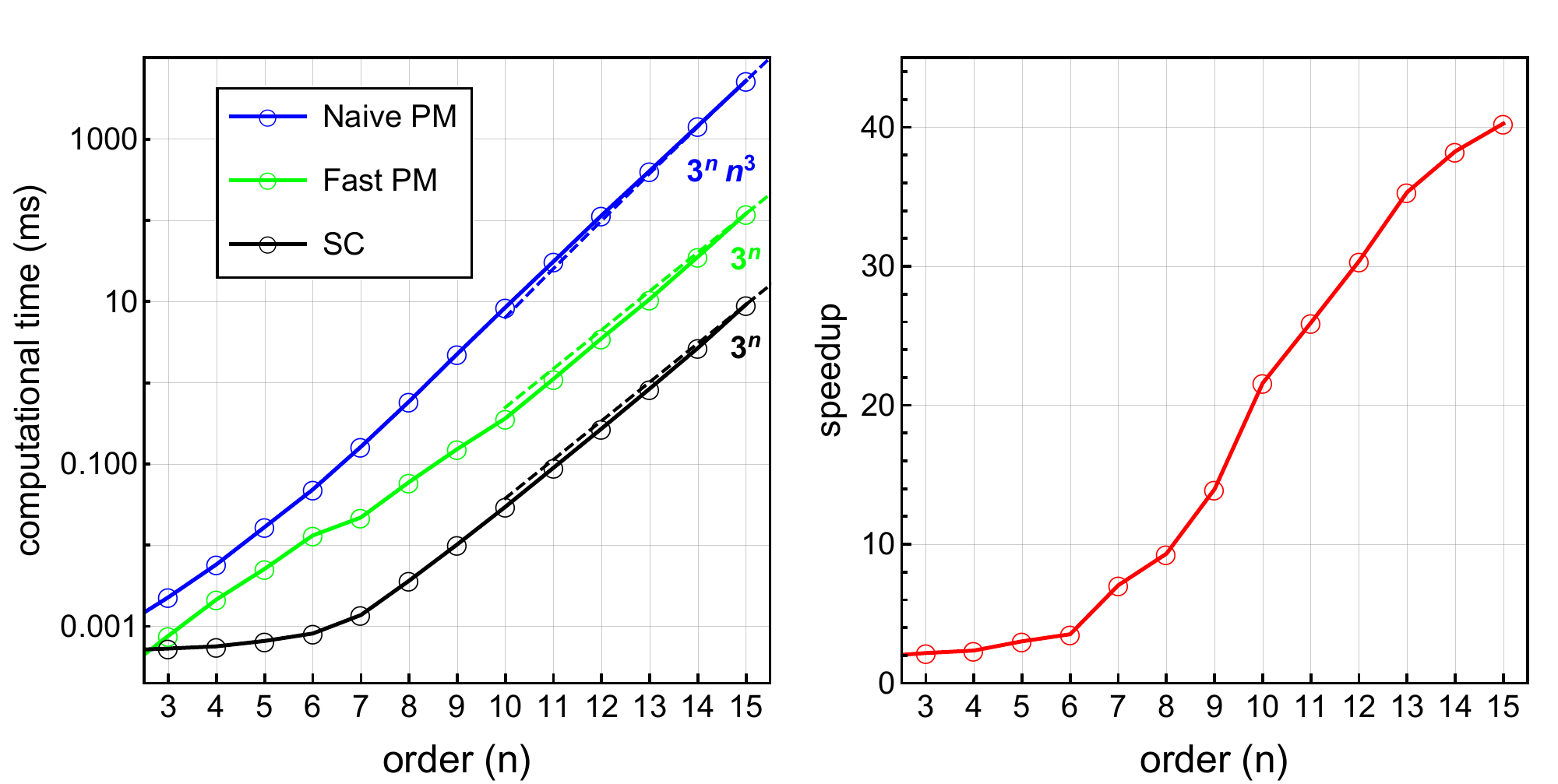}
\caption{\emph{Left}: Computational time as a function of order ($n$) for the fast principal minor (PM) algorithm within CDet symmetrized over $2^n$ vertices of two vertex types as compared to a naive determinantal implementation as well the subset convolution (SC) used by CDet. \emph{Right}: Computational speedup of the complete symmetrized CDet algorithm using fast principal minors with respect to the algorithm using a naive determinant implementation.}
\label{Fig:CDet4}
\end{figure}

\subsection{CDet for expansions with two vertex types} \label{cdet_2v}

Another possible generalization of the CDet algorithm is to expansions with two (or more) distinct types of vertices present in diagrams \cite{gull_jiali, hirsch1984charge, kanamori}. When multiple vertices are present, one way is to pick the type of each vertex insertion in diagrams stochastically. Another option is to sum over all $2^n$ possible combinations of $n$ vertices with the restriction that each vertex can only correspond to one type within a given diagram. For two vertex types, this corresponds to computing all principal minors of a $2n\!\times\!2n$ sized matrix, as described by the NAND case of the previous section. In Fig.\ref{Fig:CDet4} we show the fast and naive (determinant) implementations of such a principal minor algorithm as compared to subset convolution, which again stays unaltered from it's original CDet version. We see that even the FPM implementation, despite scaling as $O(3^n)$, is an order of magnitude slower than the subset convolution. The total gain due to using the FPM algorithm grows steadily as a function of order and reaches a factor 40 at $n=15$.
We want to note that another possible application of this algorithm is to artificially generate a second vertex by choosing an arbitrary transformation: $(\vec{r}, \tau) \rightarrow ({\vec{r}}^{\,\prime}, {\tau}^\prime)$. This transformation can either exploit some underlying symmetry which is present in the model or can be chosen stochastically. This strategy can potentially lead to a dramatic reduction of the Monte Carlo variance at the small additional computational cost factor described above. We leave the exploration of this subject to further work.

\begin{figure}
\centering
\includegraphics[width=0.48\textwidth]{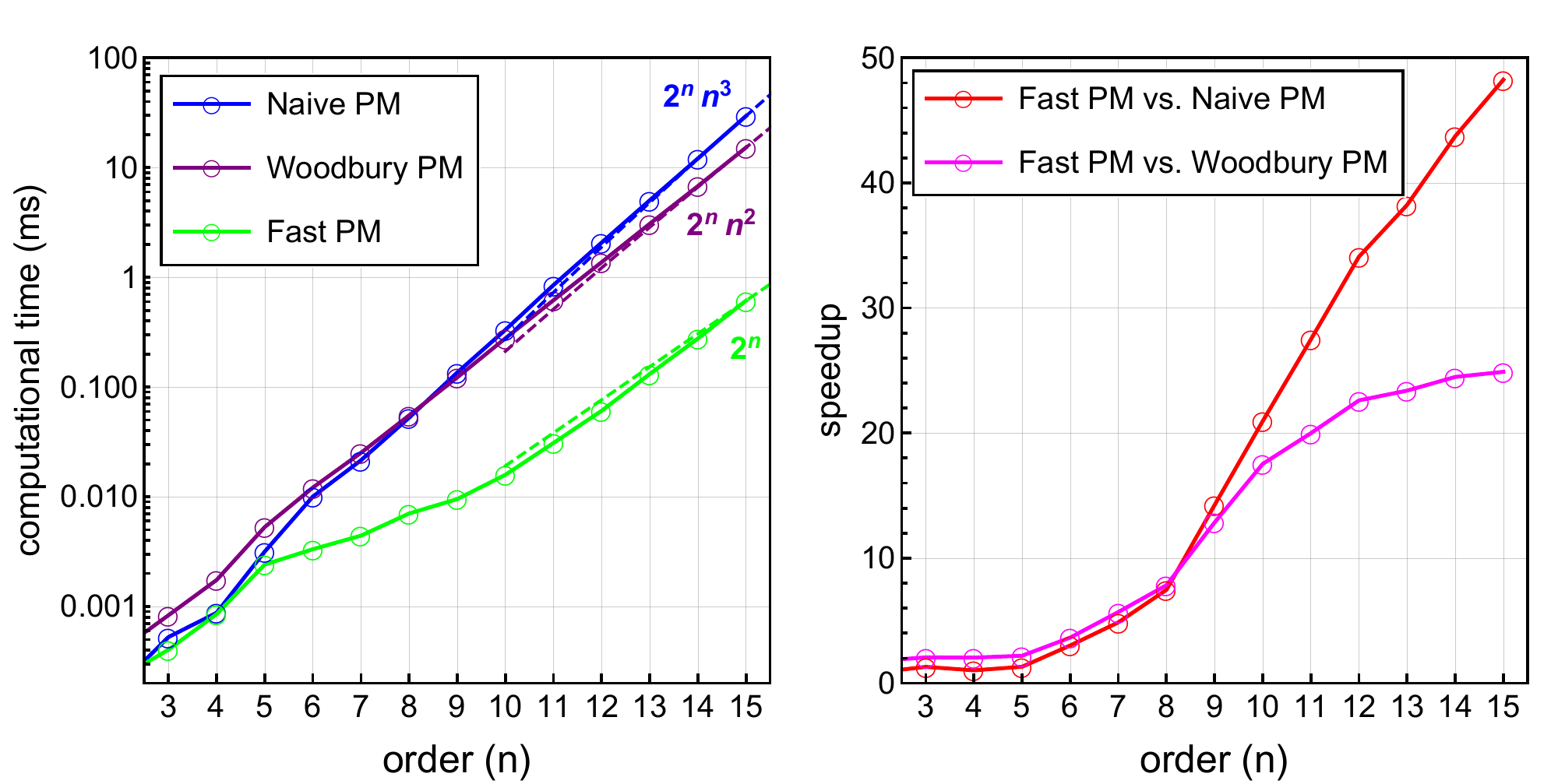}
\caption{\emph{Left}: Computational time as a function of order ($n$) for the fast principal minor algorithm within real-time
diagrammatic Monte Carlo as compared to a naive determinantal implementation as well as an implementation using Sherman–Morrison–Woodbury-type updates and Gray code. \emph{Right}: Computational speedup of the real-time diagrammatic Monte Carlo algorithm using fast principal minors with respect to the other two implementations.}
\label{Fig:CDet5}
\end{figure}

\subsection{Real-time diagrammatic Monte Carlo}

There also exists a class of diagrammatic Monte Carlo algorithms formulated directly in real-time \cite{olivier, corentin, moutenet2019cancellation, corentin2, QQMC, QQMC2} (as opposed to imaginary-time for CDet) which are based on the Keldysh formalism \cite{keldysh1965diagram}. These have the advantage that a summation over all Keldysh indices guarantees the connectivity of Feynman diagrams and therefore no recursive formula needs to be applied in order to compute connected quantities. This means that at order $n$ it is sufficient to evaluate the sum of $2^n$ determinants of
$n \! \times \! n$ matrices with elements $M^{\sigma}_{i,j} = G^{\alpha_i \alpha_j}_{0 \sigma}(X_i-X_j)$,
where $\{ \alpha_1, \ldots, \alpha_n \}$ is the set of Keldysh indices. A naive implementation
of this sum would require an effort $O(n^3 2^n)$. In current state-of-the-art implementations,
this can be brought down to $O(n^2 2^n)$ by using Sherman–Morrison–Woodbury-type updates during a Gray
code enumeration of all Keldysh indices. Let us note that while the latter more efficient approach works in most of the cases, it can also sometimes be unstable if (almost) singular matrices are generated during the update process.

A different way to look at this algorithm is to think of the sum over all Keldysh
indices as a sum of a well-defined subset of principal minors of an enlarged
$2n\!\times\!2n$ matrix built from $2\!\times\!2$ block matrices $\tilde{M}_{i,j}$
given by:
\begin{align}
  \tilde{M}_{i,j} =
  \begin{pmatrix}
  G_{0}^{++}(X_{ij}) & G_{0}^{+-}(X_{ij}) \\
  G_{0}^{-+}(X_{ij}) & G_{0}^{--}(X_{ij})
  \end{pmatrix},
\end{align}
with $i,j \in \{1, \ldots, n\}$ and $X_{ij} \equiv X_i-X_j$.
The sum over all Keldysh indices is then nothing but the sum over all 
$2^n-1$ principal minors corresponding to $n\!\times\!n$ sized sub-matrices described by the XOR case above and can be achieved in $O(2^n)$.
It should be noted that, similar to the case of computing the determinant via Schur complementation, lower-order diagrams are being computed as a side-product by this algorithm and could
potentially be used to improve the Monte Carlo variance by using, e.g. the 
Many-Configurations Monte Carlo algorithm of Ref.\cite{MCMCMC} as  well as  the  use  of  conformal  maps  applied  to  coefficients  at each  Monte  Carlo  step \cite{kim2020homotopic}. They could also
be used to compute one-particle irreducible quantities, such as the self-energy,
where a recursion involving all principal minors must be computed.


In Fig.\ref{Fig:CDet5}, we compare the scaling of the FPM against the naive (determinant) and Sherman-Morrison-Woodbury-type algorithms. We observe a steadily increasing speedup as a function of order, reaching about a factor $50$ and $25$, respectively, at order $n=15$.

\begin{figure}
\centering
\includegraphics[width=0.48\textwidth]{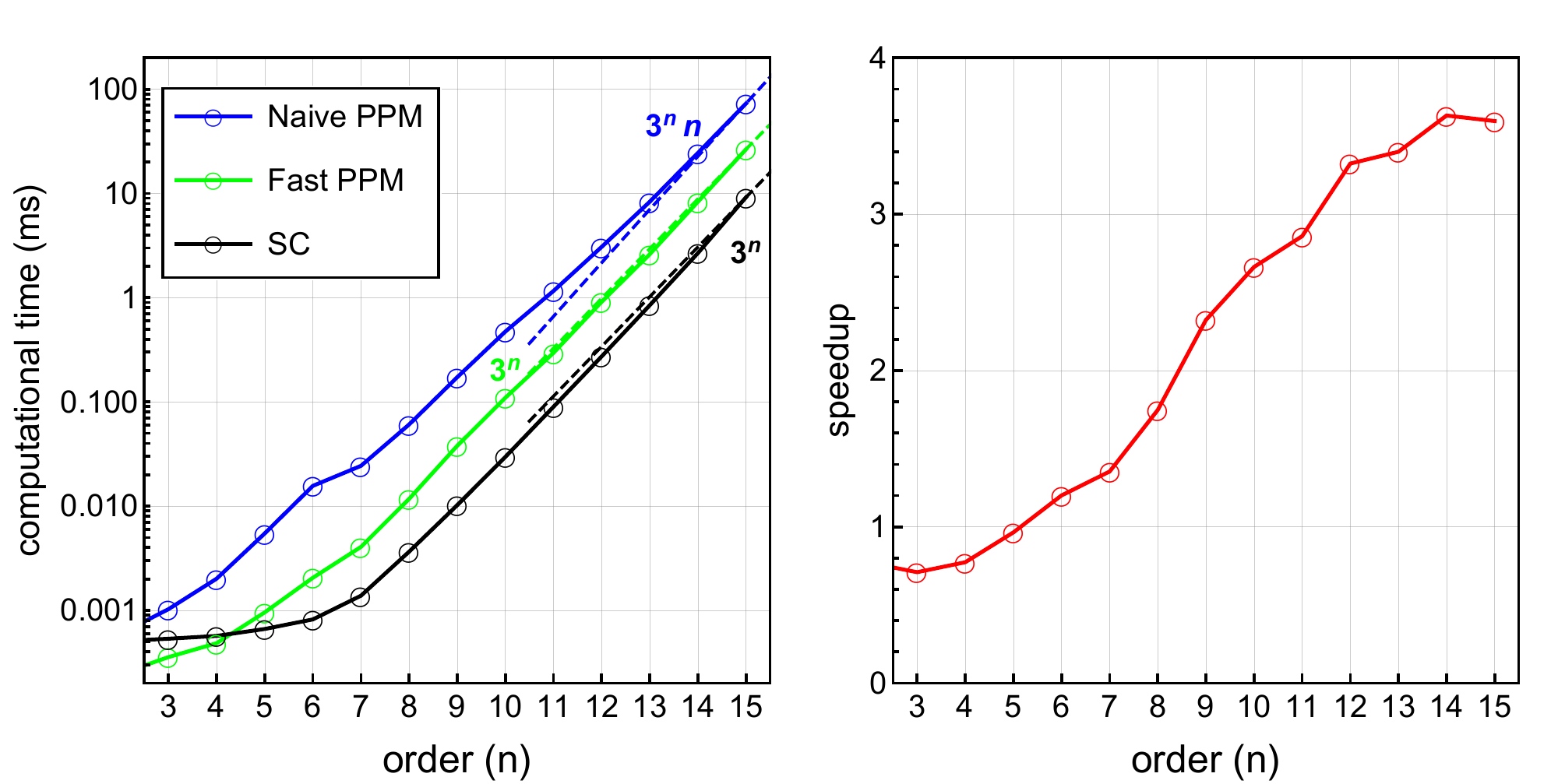}
\caption{\emph{Left}: Computational time as a function of order ($n$) for the fast permanent principal minor (PPM) algorithm as compared to a naive Ryser implementation as well the subset convolution (SC) used by CDet. \emph{Right}: Computational speedup of the complete fermionic CDet with respect to the complete bosonic CDet, both using the fastest available principal minor algorithms.}
\label{Fig:CDet3}
\end{figure}

\subsection{CTINT, DDMC and PDet}

In the context of the CTINT \cite{rubtsov2005continuous} and DDMC \cite{DDMC} algorithms, it has been shown that for fermionic systems, such as the Fermi-Hubbard model \cite{hubbard63}, all possible (connected and disconnected) Feynman diagram graph topologies contributing to the partition function can be generated as the product of two determinants (one per spin-type: $\{ \uparrow, \downarrow\}$) of matrices with entries corresponding to bare Green's functions.
\begin{align}
  a_n \sim  \int_{X_1\dots X_n} \det(M_n^{\uparrow}(X_1\dots X_n)) \det(M_n^{\downarrow}(X_1\dots X_n))
  \label{CTINT}
\end{align}
where $\left(M_n^{\sigma}\right)_{i,j} = G_{0 \sigma}(X_i-X_j)$. The evaluation of these two determinants constitutes the computational bottleneck of these algorithms, which can however be accelerated to ($O(n^2)$) by Sherman–Morrison–Woodbury-type updates \cite{sherman1950adjustment, gull_submatrix_updates} when only the position of a single vertex of the Feynman diagrams is altered at each Monte Carlo step (corresponding to the alteration of only one row and column in the matrices). An interesting observation with respect to this work is that one can use successive Schur-complementation to compute the determinants from Eq.\ref{CTINT} with the same polynomial ($O(n^3)$) complexity as Gaussian elimination or similar. The advantage is that one computes one (leading) principal minor for each order $k<n$ as a side-product. This allows for the simultaneous evaluation of all orders up to $n$ and the use of the recently introduced Many-Configurations Markov Chain Monte Carlo \cite{MCMCMC} as well as the use of conformal maps applied to coefficients at each Monte Carlo step \cite{kim2020homotopic}. It should be noted that a disadvantage of using Schur complementation to compute determinants is the lack of possibility to correct numerically-unstable pivot values without significant additional computational cost.

Another related determinantal diagrammatic Monte Carlo algorithm (PDet) has been proposed for the Fermi polaron problem in Ref.\cite{pdet}. In this case all diagrams are generated by the multiplication of a determinant in one spin with a connected set of propagators called ``backbone'' in the other, which ensures the overall connectedness of diagrams. For such algorithms, one might also benefit from the simultaneous evaluation of diagrams at all expansion orders up to $n$ by means of a Schur-determinant algorithm for leading principal minors.

\subsection{Connected diagram algorithm for bosons}

It is also possible to use the recursion~\eqref{eq:recursion} for bosonic \cite{BoseHubbard1,BoseHubbard2,BoseHubbard3,BoseHubbard4, pollet2012recent} Hamiltonians and Fermi-Bose mixtures \cite{FermiBose1,FermiBose2,FermiBose3}. Despite this natural generalization, no such calculations have been attempted to date. Here, we merely aim at discussing the computational complexity that is required to eliminate disconnected diagrams in a bosonic system. Notably, we can make use of the fast permanent principal minor algorithm introduced in the previous section. This algorithm scales as $O(3^n)$, similarly to the subset convolution in CDet, but with a slightly worse prefactor. In Fig.\ref{Fig:CDet3}, we compare the resulting computational times for fermionic and bosonic algorithms and find that the fermionic version is about $1.5$ to $3.5$ faster than the bosonic one. We believe this is negligible when compared to other factors that influence the variance of both algorithms. We leave a thorough study of bosonic system with such an algorithm to future
work.

\section{Conclusions} \label{conclusions}

In this paper we have introduced an efficient implementation of the Fast Principal Minor algorithm as well as multiple generalizations with specific applications within diagrammatic Monte Carlo
algorithms in mind. We have reported significant speedups for CDet in the normal as well as superfluid phases, and for diagrammatic schemes with two vertex types. The latter can potentially lead to reductions in the Monte Carlo variance of CDet computations. We have also introduced a generalization which reduces the asymptotic computational scaling of real-time diagrammatic Monte Carlo algorithms to $O(2^n)$ and leads to significant efficiency gains at reasonably attainable expansion orders. Finally, we have shown that CDet for bosonic systems has the same asymptotic scaling of $O(3^n)$ as for fermionic ones, despite the fact that exponentially scaling permanent principal minor computations are necessary. This opens the door for multiple applications of CDet to bosonic systems and Fermi-Bose mixtures. All of the presented algorithms can also be expanded to matrices with any fields as entries, notably truncated polynomials \cite{wietek} and nilpotent polynomials \cite{rr_sigma, rdet}. 

\section*{Acknowledgements} \label{acknowledgements}
The authors thank C.~Bertrand, R.~Costa Barroso, Y.~Fern\`andez, A.~Kim, E.~Kozik, O.~Parcollet, R.~Rossi, G.~Spada, M.~Tsatsomeros, X.~ Waintal and F.~Werner for valuable discussions. This work was granted access to the HPC resources of TGCC and IDRIS under the allocations A0090510609 attributed by GENCI (Grand Equipement National de Calcul Intensif). This work has been supported by the Simons Foundation within the Many Electron Collaboration framework.

\bibliographystyle{ieeetr}
\bibliography{main_biblio}

\end{document}